\documentstyle[amsfonts,aps]{revtex}
\draft

\begin{document}
\title{Thompson Scattering in an Expanding Universe}
\author{James L. Anderson}
\address{Department of Physics, Stevens Institute of Technology, Hoboken, NJ 07030}
\maketitle

\begin{abstract}
The Thompson cross section for scattering of electromagnetic waves by a free
electron in an expanding universe is derived here. The equations of motion
of the electron are obtained from the Einstein-Maxwell field equations of
general relativity using the Einstein-Infeld-Hoffmann surface integral
method. These integrals are evaluated approximately by perturbing off an
Einstein-deSitter cosmological field. It is found that the Thompson cross
section varies as the inverse square of the cosmic scale factor $R(t)$.
\end{abstract}


The question of whether or not the constants of nature actually vary on a
cosmological time scale was first investigated seriously by Dirac.\cite
{Dirac1} To satisfy his Large Number hypothesis, Dirac assumed that it was
the gravitational constant $G$ that varied on a cosmological time scale.
Later \cite{Dirac2} he developed this idea further by introducing the notion
of gravitational and atomic time and assumed that these two times were not
synchronous with each other.\cite{vc} However, in a recent work \cite{jla1}
(referred to hereafter as A1), I showed how one could construct model
gravitational and electromagnetic clocks whose dynamics in an expanding
universe follow directly from the Einstein-Maxwell field equations of
general relativity using the Einstein-Infeld-Hoffmann (EIH) \cite{EIH}
surface integral method without other assumptions. It was found that, for a
class of clocks for which $T_L\ll T_C\ll T_H$ where $T_L$ is the light
travel time across the clock, $T_C$ is the clock period and $T_H$ is the
Hubble time, that, to an accuracy of {\it O}$(\varepsilon ^2\delta )$ where $%
\varepsilon =T_L/T_C$ and $\delta =T_L/T_H$, the two clocks remain
synchronous with each other and measure the same cosmic time $t$ when the
cosmic gravitational field $g_{\mu \nu }$ has the form 
\begin{equation}
g_{\mu \nu }=\text{ diag}(1,-R^2,-R^2,-R^2).  \label{1}
\end{equation}
In this paper I address a different problem associated with physical
processes in an expanding universe, namely Thompson scattering. I will show
that the Thompson scattering cross section is in fact not a constant but
varies with time by a factor proportional to $1/R^2(t)$.

At first glance it might appear to the reader paradoxical that gravitational
effects associated with the expanding universe could have any effect on what
is seemingly a purely electrodynamic process. But in fact Thompson
scattering is not a purely electromagnetic process. The inertial force term
in the equations of motion for the scattering electron is of gravitational
origin as EIH first showed and, in the case of an expanding universe,
involves the scale factor $R(t)$. As in A1 the exact dependence of the
inertial force on $R(t)$ as well as the form of the electromagnetic force
due to the incident wave can be determined from the EIH surface integrals by
perturbing the gravitational and electromagnetic fields off the
Einstein-deSitter field as in A1. However, since one is dealing here with
radiation fields it turns out to be more convenient to employ conformal
coordinates $(\tau ,x,y,z)$ in which this field has the form 
\begin{equation}
g_{\mu \nu }=R^2(\tau )\text{diag}(1,-1,-1,-1)  \label{2}
\end{equation}
and where $\tau $ and $t$ are related by $dt=R(\tau )d\tau $.

The EIH surface integrals can most easily be derived from the Landau and
Lifshitz \cite{LL} form of the field equations for $g_{\mu \nu }$ and the
electromagnetic field $F_{\mu \nu }$ and have the form (in what follows I
will use units in which $G=c=1$, latin indices run from 1 to 3, greek
indices run from 0 to 3, and I employ both the Einstein summation convention
and the comma notation to denote partial derivatives) 
\begin{equation}
U^{\mu \nu \rho }=\Theta ^{\mu \nu }  \label{3}
\end{equation}
where 
\begin{equation}
U^{\mu \nu \rho }=-U^{\mu \rho \nu }=\frac 1{16\pi }\left\{ {\frak g}^{\mu
\nu }{\frak g}^{\rho \sigma }-{\frak g}^{\mu \rho }{\frak g}^{\nu \sigma
}\right\} ,_\sigma  \label{4}
\end{equation}
and 
\begin{equation}
{\frak F}^{\mu \nu }{},_\nu =0\text{ \qquad and\qquad }F_{\left| \mu \nu
,\rho \right| }=0.  \label{6}
\end{equation}
In these equations $g=$ det$(g_{\mu \nu })$, ${\frak g}^{\mu \nu }=\sqrt{-g}%
g^{\mu \nu },$ ${\frak F}^{\mu \nu }=\sqrt{-g}F^{\mu \nu },{\frak t}%
_{LL}^{\mu \nu }$ is the Landau-Lifshitz energy-stress pseudotensor and $%
T^{\mu \nu }$ is the electromagnetic energy-stress tensor given by 
\begin{equation}
T^{\mu \nu }=\frac 1{16\pi }\left( g^{\mu \nu }F_{\rho \sigma }F^{\rho
\sigma }-4g_{\rho \sigma }F^{\mu \rho }F^{\nu \sigma }\right) .  \label{7}
\end{equation}
Note that, in our expression for $\Theta ^{\mu \nu }$ above, there is no
matter contribution from the sources since they are assumed to be compact
and to vanish on and outside the EIH surfaces. This feature of the EIH
method thereby avoids having to make specific assumptions about the form of
the matter energy-stress tensor or the need to introduce singular source
terms.

Because of the antisymmetry of $F^{\mu \nu }$ and $U^{\mu \nu \rho }$ in
their indices, integration of Eq. (\ref{3}) over a closed 2-surface in a $t=$
const. hypersurface gives 
\begin{equation}
\oint \left( U^{\mu r0},_0-\Theta ^{\mu r}\right) n_rdS=0  \label{8}
\end{equation}
where $n_r$ is a unit surface normal. In a similar way one gets from Eq. (%
\ref{6}) the result 
\begin{equation}
\oint {\frak F}^{r0}{},_0n_rdS=0.  \label{9}
\end{equation}
It is these last two equations that yield equations of motion for the
sources of the gravitational and electromagnetic fields.

When the surfaces over which the integrals in Eqs. (\ref{8}) and (\ref{9})
are taken surround a source, the requirement that the surface independent
contributions vanish yield these equations. (The surface dependent terms
will in all cases vanish either identically or as a consequence of the field
equations.\cite{JNG}) These equations will be used to derive equations of
motion for a compact charge in the presence of an incident electromagnetic
field whose characteristic length scale is large compared to the size of the
EIH surfaces needed to enclose the charge. The near electromagnetic field
produced by the motion of this charge is then matched to a far radiation
field. The total flux of this scattered field is finally divided by the
incident flux to yield an expression for the Thompson scattering cross
section.

To evaluate the fields appearing in Eqs. (\ref{8}) and (\ref{9}) I follow
the methods used in A1 except that here the gravitational field is perturbed
off the background field (\ref{2}) rather than the field (\ref{1}). For
convenience the total field is written as 
\begin{equation}
{\frak g}^{\mu \nu }=R^2\widetilde{{\frak g}}^{\mu \nu }  \label{10}
\end{equation}
and $\widetilde{{\frak g}}^{\mu \nu }$as well as the electromagnetic
four-potential $A^\mu $ is expanded in a double series in $\varepsilon $ and 
$\delta .$ These fields in addition are assumed to depend on the conformal
time $\tau $ through their dependence on $\varepsilon \tau $ and $\delta
\tau $ while $R$ is a function only of $\delta t$. (In higher orders of
approximation they will of course depend on higher order multiple times.) In
addition, charges and masses are scaled so as to be {\it O}($\varepsilon ^2$%
). In what follows we will need an accuracy of $\varepsilon ^2$ and $%
\varepsilon \delta $ since we are only concerned here with the modifications
in the Newtonian dynamics in the background cosmological field.

The lowest order correction to the gravitational field is taken to be \cite
{jla2} 
\begin{equation}
\widetilde{{\frak g}}^{00}=1+\varepsilon ^2h  \label{11}
\end{equation}
and satisfies 
\begin{equation}
\nabla ^2h=0.  \label{12}
\end{equation}
For compact spherical sources (and Schwarzschild black holes) the solution
in the weak-field zone on and outside the EIH surfaces has the form 
\begin{equation}
h=4\sum \frac{\widetilde{m}_A}{r_A}  \label{13}
\end{equation}
where the index $A$ labels the sources in the system and the sum is over all 
$A$. The $\widetilde{m}_A$are as yet to be determined functions of $%
\varepsilon \tau $ and $\delta \tau $ and ${\bf r}_A={\bf x-x}_A$ where the $%
{\bf x}_A^r$ are the coordinates of the $A$th particle and are also
functions of $\varepsilon \tau $ and $\delta \tau $. When the surface
integrals in Eq. \ref{8} with $\mu =0$ are evaluated using this field one
finds that 
\begin{equation}
\partial _{\varepsilon \tau }\widetilde{m}_A=0\text{\qquad and\qquad }%
R\partial _{\delta \tau }\widetilde{m}_A=-\partial _{\delta \tau }R\,%
\widetilde{m}_A  \label{14}
\end{equation}
so that, to this order of accuracy, 
\begin{equation}
\widetilde{m}_A=\frac{m_A}R  \label{15}
\end{equation}
where the $m_A$ are constants. In a like manner one constructs the lowest
order contribution to the scalar potential 
\begin{equation}
A^0=\varepsilon ^2\phi  \label{16}
\end{equation}
with $\phi $ satisfying 
\begin{equation}
\nabla ^2\phi =0.  \label{17}
\end{equation}
When the spherically symmetric solution 
\begin{equation}
\phi =\sum \frac{\widetilde{q}_A}{r_A}  \label{18}
\end{equation}
is substituted into the surface integral (9) one obtains the result that 
\begin{equation}
\partial _{\varepsilon \tau }\widetilde{q}_A=\partial _{\delta \tau }%
\widetilde{q}_A=0  \label{19}
\end{equation}
so that 
\begin{equation}
\widetilde{q}_A=q_A  \label{20}
\end{equation}
where the $q_A$ are constants.

To derive equations of motion from the surface integrals (\ref{8}) we need
the lowest order corrections to $\widetilde{{\frak g}}^{0r}$ which are {\it %
O(}$\varepsilon ^3)$ and {\it O}($\varepsilon ^2\delta $) so we set 
\begin{equation}
\widetilde{{\frak g}}^{0r}=\varepsilon ^3h_\varepsilon ^r+\varepsilon
^2\delta \,h_\delta ^r.  \label{21}
\end{equation}
There will of course be additional corrections in higher orders of
approximation that are small compared to the first correction but large
compared to the second, e.g. post-Newtonian corrections of order $%
\varepsilon ^5$, but here we are only interested in the first order effects
of the expanding universe on Newtonian physics. As in A1 $h_\varepsilon ^r$
and $h_\delta ^r$ are determined to be 
\begin{equation}
h_\varepsilon ^r=4\sum \frac{\widetilde{m}_A}{r_A}x_A^r,_{\varepsilon \tau
}\qquad \text{and\qquad }h_\delta ^r=4\sum \frac{\widetilde{m}_A}{r_A}%
x_A^r,_{\delta \tau }\;.  \label{24}
\end{equation}
when the gauge conditions 
\begin{equation}
h_\varepsilon ^r,_r+\partial _{\varepsilon \tau }h=0\quad \text{and\quad }%
h_\delta ^r,_r+\sum \widetilde{m}_A\partial _{\delta \tau }\left( \frac 1{r_A%
}\right) =0  \label{25}
\end{equation}
are employed.

Since my purpose here is to examine Thompson scattering in an expanding
universe I will, in what follows, confine my attention to a single charged
source and take the electromagnetic field used to evaluate $\Theta ^{\mu \nu
}$ in Eq. (\ref{8}) to be a plane wave. Maxwell's equations (\ref{6}) have
the solution 
\begin{equation}
{\frak F}^{0s}=\delta _3^sE_0e^{i(\omega \varepsilon \tau -kx)}  \label{29}
\end{equation}
for a slowly varying wave propagating in the x-direction with $(\omega
\varepsilon )^2-k^2=0.$ For a wave whose spatial variation is large compared
to the size of the EIH surface surrounding the charge the surface integral
equation (\ref{8}) yields the equation of motion 
\begin{equation}
m_1x_1^r,_{\varepsilon \tau \varepsilon \tau }=-\frac 1RqE_0e^{i\omega
\varepsilon \tau }  \label{30}
\end{equation}
which has the solution 
\begin{equation}
x_1^3=\frac 1R\frac{q_1}{m_1}\frac 1{\omega ^2}E_0e^{i\omega \varepsilon
\tau }.  \label{31}
\end{equation}

It remains finally to compute the scattered field produced by the motion of
our charge. One possible way to do this would be to introduce a model source
term into the right hand side of the first of Eqs. (\ref{6}) whose motion is
characterized by $x_1^3$ above. However, in keeping with the EIH philosophy
of not specifically introducing sources into the field equations and because
the EIH procedure makes it unnecessary, we will proceed by constructing a
radiation solution whose inner expansion matches on to the outer expansion
of the near field $\phi $ in Eq. (\ref{18}) with $r_A$ computed using $x_1^3$
above. This outer expansion in inverse powers of $r$ is given by 
\[
\phi =\frac{q_1}{\left( r^2+r_1^2-2rr_1\cos \theta \right) ^{1/2}} 
\]
\begin{equation}
=\frac{q_1}r+\frac{q_1}{r^2}r_1\cos \theta +O(1/r^3).  \label{32}
\end{equation}
The corresponding outer dipole field is given by 
\begin{equation}
\phi =\frac a{\varepsilon r}+\left\{ \frac{W^{\prime }(\varepsilon (\tau -r))%
}{\varepsilon r}+\frac{W(\varepsilon (\tau -r))}{(\varepsilon r)^2}\right\}
\cos \theta  \label{33}
\end{equation}
whose inner expansion is 
\begin{equation}
\phi =\frac a{\varepsilon r}+\frac{W(\varepsilon \tau )}{(\varepsilon r)^2}%
+O(1)  \label{34}
\end{equation}
where $a$ and $W(\varepsilon \tau )$ are to be determined by matching.
Comparing Eqs. (\ref{32}) and (\ref{34}) we see that 
\begin{equation}
a=\varepsilon q_1\qquad \text{and\qquad }W(\varepsilon \tau )=\varepsilon
^2q_1r_1.  \label{35}
\end{equation}
The gauge condition $A^\mu ,_\mu =0$ used to derive the wave equations for $%
A^\mu $ allows us to derive an expression for the vector potential part of $%
A^\mu $ corresponding to $\phi $ given by 
\begin{equation}
A_3=-\frac{\varepsilon q_1r_1,_{\varepsilon \tau }}r.  \label{36}
\end{equation}

With the above expressions for $\phi $ and $A_3$ we can now calculate $%
F_{\mu \nu }$ which in turn allows us to determine the Poynting vector $%
S^r=T^{0r}$ to be 
\begin{equation}
S^r=\frac 1{4\pi }\frac{q_1^2}{r^2}\left\{
(1+n_3)n_1,(1+n_3)n_2,-(n_1^2+n_2^2)\right\} (r_1,_{\tau \tau })^2+O(\frac 1{%
r^3}).  \label{37}
\end{equation}
where $n_r$ is a unit vector. The total integrated flux $f$ is then
calculated to be 
\begin{equation}
f=\frac 23q_1^2(r_1,_{\tau \tau })^2=\frac 23\left( \frac 1R\frac{q_1^2}{m_1}%
E_0\right) ^2.  \label{38}
\end{equation}
The Thompson scattering cross section $\sigma _\tau $, equal to $f$ divided
by the incoming flux $E_0^2/4\pi $, is thus given by 
\begin{equation}
\sigma _T=\frac{8\pi }3\left( \frac 1R\frac{q_1^2}{m_1}\right) ^2  \label{39}
\end{equation}
and is seen to differ from the usual expression for $\sigma _T$ by the
factor of $1/R^2$.

This result raises a number of complex questions. First and foremost, is it
true or does some mechanism shield elementary processes such as Thompson
scattering from the effects of the expanding universe? Do all cross
sections, e.g. nuclear, experience this same effect? If the answer to both
questions is yes, then a number of issues in stellar evolution and
nucleosynthesis in the early universe will of necessity have to be
reexamined.

\end{document}